\documentclass{ivs2e}     
\usepackage{epsfig}       
\usepackage{floatfig}     

\PaperSize{A4}

\Title{Measurement of core-shifts with astrometric multi-frequency calibration}
 
\STitle{Measurement of source's core-shifts}  

\PublicationType{proceedings} 

\PublicationName{17th European VLBI Meeting Proceedings}

\IvsComponentName{Network Center}

\NumberofAuthors{5}

\AuthorName{1}{Mar\'{\i}a Rioja}
\AuthorName{2}{Richard Dodson}
\AuthorName{3}{Richard Porcas}
\AuthorName{4}{Hiroshi Suda}
\AuthorName{5}{Francisco Colomer}

\ContactAuthorReference{1}

\AuthorEmail{1}{rioja@oan.es}
\AuthorEmail{2}{dodson@vsop.isas.ac.jp}
\AuthorEmail{3}{porcas@mpifr-bonn.mpg.de}
\AuthorEmail{4}{hiroshi.suda@nao.ac.jp}
\AuthorEmail{5}{colomer@oan.es}

\AuthorTelephone{1}{+34 91 8855060}
\AuthorTelephone{2}{}
\AuthorTelephone{3}{+49 (0)228 5250}
\AuthorTelephone{4}{+81 (0)422 343600}
\AuthorTelephone{5}{}

\AuthorPersonalWebPage{1}{}   
\AuthorPersonalWebPage{2}{}   
\AuthorPersonalWebPage{3}{}   
\AuthorPersonalWebPage{4}{}   
\AuthorPersonalWebPage{5}{}   

\AuthorInstitutionReference{1}{1}
\AuthorInstitutionReference{2}{2}
\AuthorInstitutionReference{3}{3}
\AuthorInstitutionReference{4}{4}
\AuthorInstitutionReference{5}{1}

\NumberofInstitutions{4}

\InstitutionName{1}{Observatorio Astron\'omico Nacional, Alcal\'a de 
Henares, Spain}
\InstitutionName{2}{Institute of Space and Astronautical Science, JAXA, 
Sagamihara, Japan}
\InstitutionName{3}{Max-Planck-Institut fuer Radioastronomie, Bonn, Germany}
\InstitutionName{4}{VERA Observatory, NAOJ, Mitaka, Tokyo, Japan;
Dep. Astronomy, Univ. Tokyo, Japan}

\InstitutionPostAddress{1}{Apdo. 112, 28803 Alcal\'a, Madrid}
\InstitutionPostAddress{2}{Yoshinodai, Sagamihara, Kangawa 229-8510, Japan}
\InstitutionPostAddress{3}{Auf dem Huegel,69,D-53121 Bonn, Germany}
\InstitutionPostAddress{4}{2-21-1 Osawa, Mitaka, Tokyo 181-8588, Japan}

\InstitutionCountry{1}{Spain}
\InstitutionCountry{2}{Japan}
\InstitutionCountry{3}{Germany}
\InstitutionCountry{4}{Japan}

\InstitutionFax{1}{+34 91 8855062}
\InstitutionFax{2}{+81 (0)422 343600}
\InstitutionFax{3}{+49 (0)228 525229}
\InstitutionFax{4}{+81 (0)422 343600}

\InstitutionWebPage{1}{http://www.oan.es}
\InstitutionWebPage{2}{http://www.nao.ac.jp}
\InstitutionWebPage{3}{http://www.mpifr-bonn.mpg.de}
\InstitutionWebPage{4}{http://www.nao.ac.jp}
\begin{document}
\initfloatingfigs
\MakeTitle                

\begin{abstract}

VLBI is unique, among the space geodetic techniques,  in its contribution 
to defining and maintaining the International Celestial Reference Frame,
providing precise measurements of coordinates of extragalactic 
radiosources. 
The quest for increasing accuracy of VLBI geodetic products 
has lead to a deeper revision of all aspects that might introduce errors 
in the analysis.
The departure of the observed sources  from perfect, stable, compact and 
achromatic celestial targets falls within this category. This paper is 
concerned with the impact of unaccounted frequency-dependent position 
shifts of source cores in the analysis of dual-band S/X VLBI geodesy 
observations, and proposes a new method to measure them. The multi-frequency 
phase transfer technique developed and demonstrated by Middelberg 
et al. (2005) increases the high frequency coherence times of VLBI 
observations, using the observations at a lower frequency.
Our proposed {\sc Source/frequency phase referencing} method endows it with 
astrometric applications by adding a strategy to estimate the ionospheric 
contributions. Here we report on the first successful application
to measure the core shift of the quasar 1038+528 A at 
S and X-bands, and validate the results by comparison
with those from standard phase referencing techniques. 
In this particular case, and in general in the cm-wavelength regime, 
both methods are equivalent.  
Moreover the proposed method opens a new horizon with targets and fields 
suitable for high precision as\-tro\-me\-tric studies with VLBI, especially at 
high frequencies where severe limitations imposed by the rapid
fluctuations in the troposphere prevent the use of standard phase 
referencing techniques. 

\end{abstract}

\section {Introduction}

\noindent
Geodetic VLBI observations with a network of antennas at the Earth 
are affected by the pro\-pa\-ga\-tion medium, mainly the ionosphere and the 
troposphere. 
It is a basic practice in geodesy to calibrate the ionospheric contribution 
with simultaneous observations at S/X-bands (2.2GHz/8.4GHz, respectively).
The ionospheric-free delay observables at 
X-band ($\tau_x^{c}$) are estimated from a combination of observed 
delays ($\tau_x, \, \tau_s$) at both bands: 

\begin{center}
$\tau_x^{c} \, = \, {\nu_x^2 \over  {\nu_x^2 - \nu_s^2}} 
\, . \, \tau_x - {\nu_s^2 \over  {\nu_x^2 - \nu_s^2}} \,.\, \tau_s$ \\
\end{center}

\noindent
and used to estimate the geodetic parameters. 
This approach works under the critical assumption 
that the brightness distributions for each source are identical 
and are co-located at both frequencies. 
In sources for which the VLBI core position is frequency dependent
the exact expression must include a 24-hour sinusoidal extra term 
whose amplitude depends on the magnitude of the shift of the position 
of the source core ({\it core shift}): 

\begin{center}
$\, + \, {\nu_s^2 \over  {\nu_x^2 - \nu_s^2}} \, \ast \, \Delta 
\tau_{sx}^{geo} \sim 0.08 \, \ast \, \Delta \tau_{sx}^{geo} $ \\
\end{center}

\noindent
where $\Delta \tau_{sx}^{geo} = {\vec{D} \,.\, \vec{\theta}_{sx} 
\over c}$, and   $\vec{\theta}_{sx}$ corresponds to the 
{\it core shift} between S and X-bands. \\ 

\noindent
The non-inclusion of this extra term introduces errors in the estimated 
ionosphere-free observables, and hence on the astrometric/geodetic products 
from the analysis. For sources with non-varying (i.e. stable) {\it core shift} 
$\theta_{sx}$, the unaccounted extra term will propagate into an offset 
from the true X-band position, of magnitude ($\theta_{sx} 
\, \ast \, 0.08$) in the direction away from the S-band position 
\cite{porcas95}. Instead, unstable {\it core shifts} can propagate also 
into uncertainties in the estimated Earth orientation parameters
(Engelhardt, these proceedings) in the multi-epoch geodetic analysis.
Section 2 is concerned with the origin, magnitude and geodetic impact of
core shifts; Section 3 discusses the methods to measure them and Section 4
presents the results of our proposed method.

\section{Core shifts do exist}

\noindent
Changes in the observed core positions at different frequencies 
have been measured in several sources, for example, 1038+528 A 
\cite{marcaide84},\cite{rioja00}, 4C39.25 \cite{guirado95},
3C395 \cite{lara96} and
1823+568 \cite{paragi00}.
\noindent
We propose to classify the core shifts in two groups depending on their 
origin: \\

\noindent
$\bullet$ ``Astronomical core shifts'' result from 
opacity effects in the jet. The unresolved ``core'' of a compact 
extragalactic radio source is believed to mark the location where the 
optical depth to synchrotron self absorption $\sim  1$. 
This position changes with observing frequency as 
$R_{core} \propto \nu^{-1/k_r}$, where $k_r$ depends on physical 
conditions in the jet. Core position shifts between 
S and X-bands 
of up to 1.5 milli-arcsecond ($mas$) are predicted in \cite{lobanov98}.\\

\noindent
$\bullet$ ``Instrumental core shifts'' result from convolving
the source structure with different resolutions at different frequencies,
causing core shifts of up to half the beam size at the 
lower frequency. \\

\noindent
While the existence of source ``{\it core shifts}'' cannot be predicted, 
there are some clues which alert one to them. 
Larger ``astronomical core shifts'' are expected for flat spectrum sources, 
where the power index $k_r \, \sim 1$; 
``instrumental core shifts'' can be expected if the source structure at 
the higher frequency falls within a small fraction of the beam size at 
the lower frequency. Regardless of its nature, both core shifts have 
an identical 
effect on the analysis of S/X geodesy data. Table 1 lists the propagation of
plausible unaccounted stable core shifts into the analysis products.

\begin{center}
\begin{tabular}{c|cc|c}
\hline
Observing      & ``Astronomical'' & ``Instrumental'' & Source position \\ 
frequencies    &    core shift    & core shift & error \\ \hline \hline
2.2/8.4 GHz & $0-1\, mas$ & $0-2\, mas$ & $0-200 \,\mu as$ \\ 
8.4/30 GHz & $0-0.25 \,mas$ & $0-0.5\, mas$ & $0-50\,\mu as$ \\ \hline
\end {tabular}
\end{center}
\noindent
Table 1: Propagation of unaccounted non-varying core shifts into an offset
from the true X and K-band position in the geodetic analysis of 2.2/8.4 GHz 
(S/X) and 8.4/30 GHz (X/K) observations, respectively. \\

\begin{figure}[!h]
\centerline{\psfig{figure=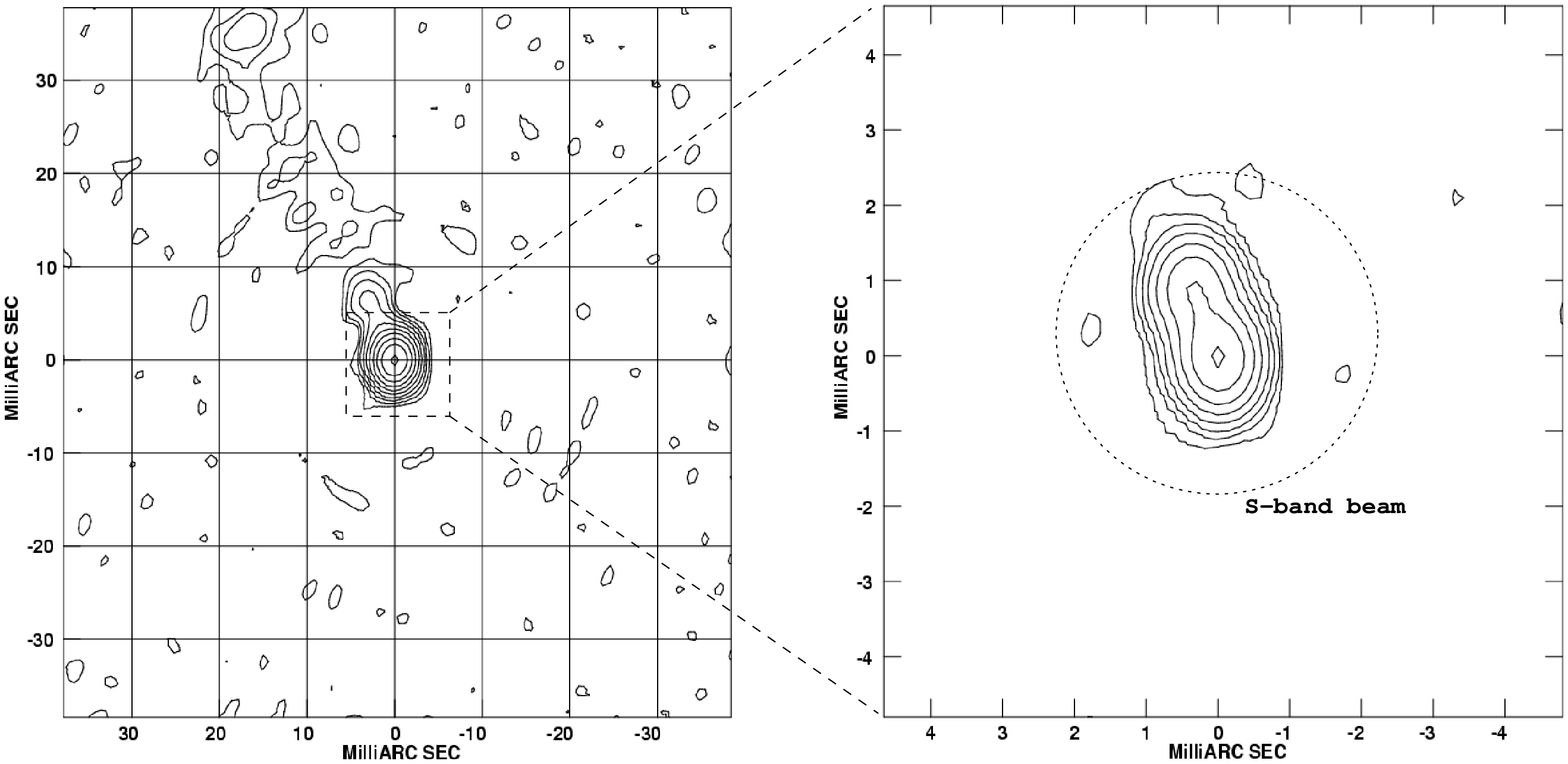,angle=-90,width=9cm} 
}
  \caption{Hybrid map of 1038+528 A at S-band ({\it left}),
and at X-band ({\it right}).
The S-band beam superimposed on the X-band map illustrates the case of
structure blending effects from insufficient resolution at lower
frequencies, and therefore ``instrumental'' core shifts.} 
  \label{sigma}
\end{figure}

\section{Ways to measure frequency-dependent core shifts}

\noindent
Using closure phase relations in the image processing of VLBI data results in
lack of absolute positional information in the hybrid maps. A rigorous 
alignment of maps at different frequencies, to measure frequency-dependent 
shifts of the core position, requires absolute astrometry observations,  
or standard phase-referencing to a nearby (achromatic) radio source. 
If astrometric observations are not feasible, a simpler but more imprecise
procedure for extended sources is to use an optically thin component 
to align maps at different frequencies, and close epochs, and then 
estimate the change in the position of the core. 

\noindent
Recently, Middelberg et al. \cite{enno04,enno05} proposed a new 
astrometric method that uses fast frequency switching observations of the 
target source and relies on the transfer of calibration from the lower to 
the higher frequency, after scaling by the frequency ratio. 
Their implementation  proved to be a successful strategy to calibrate the
rapid fluctuations of the troposphere, and hence extended the coherence time,
in VLBA observations at 86 GHz, using interleaved scans at 15 GHz. This
allowed the detection of a very weak, 100 mJy source. It
also served to unveil the non-integer frequency ratio problem in the 
application of this method.
On the other hand, the unaccounted dispersive ionospheric contamination, 
which was non-negligible even at these high frequencies,  
prevented them from making a proper astrometric measurement of the 
core shift. \\

\noindent
We present an extension of this method, a so-called ``{\sc Source/frequency 
phase referencing}'' which complements the fast frequency 
switching observing strategy with source switching, in order  
to calibrate the remaining dispersive contaminating contributions 
to the observables. Section 4 describes the first successful
astrometric measurement of a {\it core shift} with this method,
and Appendix A contains a brief discussion of the basics of the method.\\

\section{Source/frequency phase-referencing}

Conventional VLBI at high frequencies is severely constrained  
by the short coherence times imposed by the rapid fluctuations in the 
troposphere. 
The non-dispersive nature of the tropospheric propagation
makes it possible to use lower frequency observations
to calibrate higher, providing the switching interval between frequencies
matches the temporal structure of the tropospheric fluctuations 
(i.e. coherence time) at the lower frequency. This is the basis
of the multi-frequency phase transfer technique developed and demonstrated by
Middelberg et al. \cite{enno04,enno05}.
Our {\sc Source/frequency phase referencing} method 
adds a source switching observing strategy, in addition to the fast
frequency switching, to calibrate
the dispersive contributions in the phase transfer strategy. 
The nodding between sources has to match the temporal and spatial
structures of ionospheric pro\-pa\-ga\-tion, and other non-dispersive terms, 
such as instrumental based contributions. A complete description of 
the method is given in Appendix A. 

We have successfully applied this method to the astrometric a\-na\-ly\-sis
of dual-band S/X VLBA observations of the pair of quasars 1038+528 A and B, 
$33^ {\prime \prime}$ apart, and measured the core shift in the quasar 
1038+528 A. 
The calibration transfer between frequencies involves multiplying the phases
by the frequency ratio. The calibration transfer between 
sources is done as in standard phase referencing.
The result is a {\sc Source/frequency phase referenced} map, shown in figure 2,
whose offset from the center is a direct measure of the combined core shifts 
in the two sources between the two frequencies. The interpretation in terms
of individual contributions from each source, assuming that shifts in the 
core position in each source would occur along the jet axis directions, 
is simplified by the quasi-orthogonal structures in this pair. 
The close alignment of the core shift offset with the A quasar source axis
suggests a dominant contribution arising from this quasar; moreover, the
quantitative agreement between the magnitude of the offset and the separation
between the core and second component in the map of quasar A at X-band 
suggests the ``instrumental'' dominant nature of the offset.
The results from a previous standard phase referencing analysis \cite{rioja00}
are in complete agreement with those presented here, validating this 
new approach.

\begin{figure}[!h]
\centerline{\psfig{figure=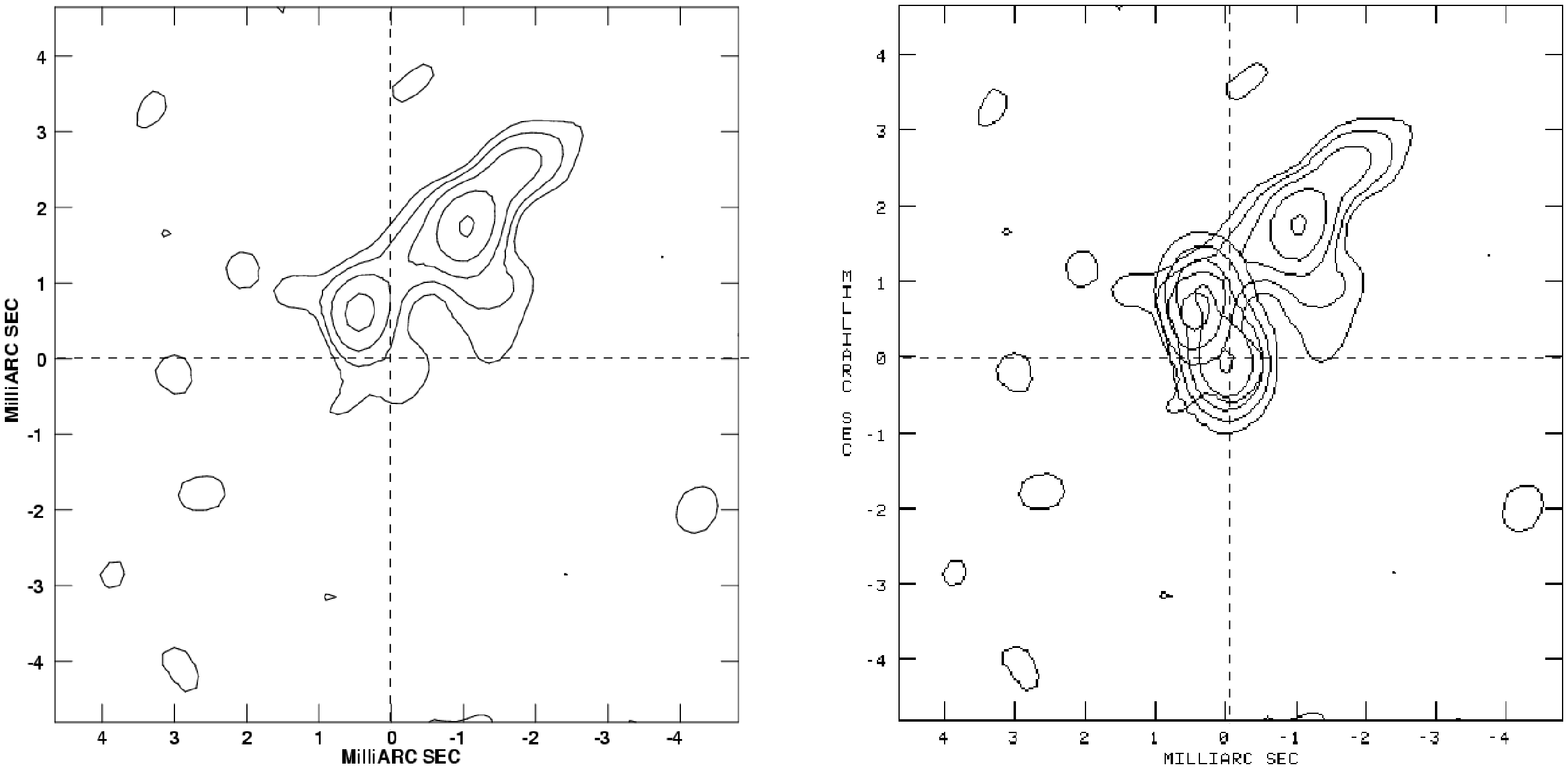,angle=-90,width=9cm} 
}
  \caption{{\it Left:} {\sc Source/frequency phase referenced} map 
of 1038+528 B from
S/X VLBA observations; the $\sim 800 \mu as$ offset in NE direction
is an estimate of the combined core shift of 1038+528 A and B quasars
between S and X bands.
{\it Right:} Same map, with the hybrid map of 1038+528 A quasar superimposed, 
to show the agreement between the offset and the separation between the 2 
components in the hybrid map of A quasar. This is an argument in favor of 
``instrumental core shift''.}
  \label{sigma}
\end{figure}

\section{Conclusions}

\noindent
Unaccounted core shifts in the analysis of dual frequency VLBI geodesy 
observations propagate into offsets from the true X-band positions
for the ICRF. We estimate deviations up to 200 $\mu as$
and 50 $\mu as$, respectively, for S/X and X/K observations, assuming 
temporally stable core shifts.   
Moreover, unstable core shifts can also corrupt the estimated Earth 
orientation parameters.

\noindent 
We have successfully applied the {\sc Source/frequency phase 
referencing}
method to the a\-na\-ly\-sis of dual band S/X VLBA
observations of the pair of quasars 1038+528 A and B, and measured a core
shift in quasar 1038+528 A of ca. 800 $\mu as$. Our result is
equivalent to, within the errors, to
those obtained using standard phase referencing techniques \cite{rioja00}.
As far as we know this is the first case of successful astrometric application
of this multi-frequency phase transfer method. 
In this case the simultaneous observations of both frequencies and sources
allowed solutions despite the non-integer ratio of the observed frequencies.  
While this new strategy does not present any advantage with respect to 
traditional techniques in the cm-wavelength regime, it does hold a big 
potential at high frequencies, which are out of the range of conventional 
phase-referencing. In particular, we foresee a big impact when applied to 
observations of molecular line emission, where it may provide bona fide
astrometric alignment of emission arising from different transitions and 
help to elucidate the controversy between the proposed pumping mechanism
for masers in evolved stars. \\

\small

\section* {Appendix A. The basics of the new method}

\noindent
This section outlines the basics of this astrometric method 
{\sc Source/frequency phase referencing} aimed to measure core shifts in 
radio sources. Its application involves observations of the target and 
a nearby source (in the formulae, $A$ and $B$) at the two frequencies 
of interest (here $x$ and $s$).
\noindent
At the post-processing, the amplitude calibration 
is done using traditional techniques, for all observations;
a pure self-calibration analysis is used to solve for the phase, delay
and rate of each low frequency ($s$) target source ($A$) observations.
Following the standard nomenclature, the phase values $\phi_A^s$ are 
shown as a compound of geometric, tropospheric, ionospheric and 
instrumental terms - assuming that structural contributions $\phi_{A,str}^s$ 
have been computed using the hybrid maps, and removed: \\

$\phi_A^s = \phi_{A,geo}^s + \phi_{A,tro}^s + \phi_{A,ion}^s 
+ \phi_{A,inst}^s + 2\pi n^s_A \,\,$, with $n^s_A$ integer \\ 

\noindent
These values are scaled by the frequency ratio, $R$, and
used to calibrate the high frequency observations. 
The resultant high frequency referenced phases to the low frequency are: \\

$\phi_A^x - R.\phi_A^s = \phi_{A,str}^x + (\phi_{A,geo}^x - R.\phi_{A,geo}^s) 
+ (\phi_{A,tro}^x - R.\phi_{A,tro}^s) + (\phi_{A,ion}^x - R.\phi_{A,ion}^s) \\
\hspace*{2.5cm} + (\phi_{A,inst}^x - R.\phi_{A,inst}^s) 
+ 2\pi (n^x_A-R.n^s_A) \hspace*{7cm} (1)$  \\ 

\noindent
This calibration strategy results in perfect cancellation of non-dispersive 
tropospheric terms:\\
$(\phi_{A,tro}^x - R.\phi_{A,tro}^s) = 0 \, $, 
but not for the dispersive ionosphere: 
$(\phi_{A,ion}^x - R.\phi_{A,ion}^s) = (R-{1 \over R}) \phi_{A,ion}^s$ \\

\newpage

\noindent
Taking this into account in Equation (1) above results in: \\

$\phi_A^x - R .\phi_A^s = \phi_{A,str}^x + 2 \pi \nu {\vec{D}\,.\, 
\vec{\theta}_{A,sx} \over c} 
+ (R-{1 \over R}) \phi_{A,ion}^s
+ (\phi_{A,inst}^x - R .\phi_{A,inst}^s) + 2\pi (n^x_A-R.n^s_A) \hspace*{1cm}
  (2)$ \\ 

\noindent
where $2 \pi \nu {\vec{D} \,.\, \vec{\theta}_{A,sx} \over c} =
(\phi_{A,geo}^x - R .\phi_{A,geo}^s)$, and $\vec{\theta}_{A,sx}$ is
the core shift in $A$ between $s$ and $x$.\\

\noindent
Similarly, the analysis of the observations of a nearby calibrator, $B$,
after removing structural terms, $\phi_{B,str}^x$ and $\phi_{B,str}^s$
at both frequencies, results in: \\

$\phi_B^x - R. \phi_B^s = 2 \pi \nu {\vec{D}.\vec{\theta}_{B,sx} \over c} 
+ (R-{1 \over R}) \phi_{B,ion}^s
+ (\phi_{B,inst}^x - R. \phi_{B,inst}^s) + 2\pi (n^x_B-R.n^s_B) 
\hspace*{2.3cm} (3)$ \\ 

\noindent
which are transfered for further calibration of $A$ source observations.
The resultant {\sc Source/frequency referenced} phases, 
combining (2) and (3), are: \\

($\phi_A^x - R. \phi_A^s) - (\phi_B^x - R .\phi_B^s) =  
 \phi_{A,str}^x + 2 \pi \nu {D. (\vec{\theta}_{A,sx}-\vec{\theta}_{B,sx}) 
\over c} 
+ (R-{1 \over R}) (\phi_{A,ion}^s - \phi_{B,ion}^s) 
+ (\phi_{A,inst}^x - R.\phi_{A,inst}^s) \\
\hspace*{4.5cm} -  (\phi_{B,inst}^x - R.\phi_{B,inst}^s)
 + 2\pi [(n^x_A-R.n^s_A) - (n^x_B-R.n^s_B)]$ \\

\noindent
A careful planning of the observations, namely switching between 
sufficiently nearby sources with a duty cycle which matches the 
ionospheric/instrumental time-scale variations, would result in
negligible differential ionospheric error and instrumental corruption, that 
is: \\

$\hspace*{1cm} (R-{1 \over R}) (\phi_{A,ion}^s - \phi_{B,ion}^s) \sim 0 $ \\
$\hspace*{1.5cm} (\phi_{A,inst}^x - R . \phi_{A,inst}^s)-  
(\phi_{B,inst}^x - R . \phi_{B,inst}^s) \sim 0 $ \\

\noindent
which results in an expression for the {\sc Source/frequency 
referenced} phases for the target source free of ionospheric/instrumental 
corruption: \\

($\phi_A^x - R.\phi_A^s) - (\phi_B^x - R.\phi_B^s) 
= \phi_{A,str}^x + 2 \pi \nu {\vec{D} . (\vec{\theta}_{A,sx}-
\vec{\theta}_{B,sx})\over c} + 2\pi n^{\prime \prime} \,$,  with
$n^{\prime \prime}$ integer if $R$ integer (or $n^s_A = n^s_B$) \\

\noindent
And finally, the calibrated complex visibilities from the target observations
are inverted to yield a synthesis image of $A$ at $x$-band, 
where the offset from the center is an estimate of the combined core shifts 
in $A$ and $B$ between $s$ and $x$-bands.\\


\begin{thebibliography}{99}

\bibitem{guirado95}
Guirado, J.C., Marcaide, J.M., Alberdi, A., et al. 1995, AJ, 110, 2586 
\bibitem{lara96}
Lara, L., Marcaide, J.M., Alberdi, A., Guirado, J.C. 1996, A\&A, 314, 672
\bibitem{lobanov98}
Lobanov, A. P. 1998, A\&A, {\bf 330}, p. 79
\bibitem{marcaide84}
Marcaide, J.M. \& Shapiro, I.I. 1984, ApJ, 276, 56 
\bibitem{enno04}
Middelberg, E., PhD Thesis, 2004 
\bibitem{enno05}
Middelberg, E., Roy, A. L., Walker, R. C., Falcke, H., 2005, A\&A, {\bf 433},
p. 897-909
\bibitem{paragi00}
Paragi, Z., Fejes, I., Frey, S., 2000, Proc. IVS General Meeting, 
p.342
\bibitem{porcas95}
Porcas, R., Patnaik, A., 1995, Proc. 10th Working Meeting on European VLBI for
Geodesy and Astrometry, p. 188 
\bibitem{rioja00}
Rioja, M.J., Porcas, R.W., 2000, A\&A, {\bf 355}, p.552-563 
\end{thebibliography}
\end{document}